\documentclass[twocolumn, amsmath, superscriptaddress, amsfonts,prb]{revtex4-2}
\usepackage{graphicx}
\usepackage{url}
\pdfoutput=1
\usepackage{mhchem}

\begin{document}

\title{The FLASH enigma}

\author{Diana Shvydka}\email{diana.shvydka@osumc.edu}\affiliation{Department of Radiation Oncology, The Ohio State University Wexner Medical Center, Columbus, OH 43221, USA}
\author{Victor Karpov}\email{victor.karpov@utoledo.edu}\affiliation{Department of Physics and Astronomy, University of Toledo, Toledo,OH 43606, USA}
\author{Nilendu Gupta}\email{Nilendu.Gupta@osumc.edu}\affiliation{Department of Radiation Oncology, The Ohio State University Wexner Medical Center, Columbus, OH 43221, USA}

\begin{abstract}
We consider physics behind the FLASH modality of cancer radiation treatment where extremely short treatment times are achieved with ultra high dose rates maintaining the conventional antitumor effectiveness and yet substantially reducing damage to normal tissues (sparing effect). The difference in responses between normal and tumor tissues is attributed here to different recombination rates related to their structure morphologies: ordered in normal vs disordered in the tumor tissues. Correspondingly different are their charge densities under ionizing radiation. In normal tissues it is high enough to form electron-hole liquid (EHL). Because of low EHL diffusivities, the chemical reaction and generation of free radicals are suppressed; hence, sparing effect. To the contrary, a disordered tumor tissue renders efficient energy relaxation channels forming antitumor free radicals. We describe the FLASH thresholds for doses and dose rates.
\end{abstract}

\maketitle

\section{Introduction}\label{sec:intro}

While radiation therapy (RT) is a powerful method for treating cancers, its potential to damage healthy tissues often limits the prescribed radiation dose, and the ability to escalate doses to tumors further to obtain better tumor control. 
The past decades of technological advances were focused on the equipment and treatment approaches providing high radiation beam conformity to the tumor target and real-time imaging, immediately before or during the RT delivery. This conventional radiotherapy (CONV-RT) typically requires multiple sessions, often spanning weeks, and relies on the classical radiobiology principles. Recently the focus has shifted to a dose rate modulation, unexpectedly providing a drastic increase in the therapeutic index with FLASH radiotherapy (FLASH-RT), leveraging average ultra-high dose rates (UHDR) of greater than 40 Gy/s, orders of magnitude higher than those of CONV-RT. A unique characteristic of FLASH-RT is that its significantly reduced damage to healthy tissues does not compromise the antitumor effectiveness of the treatment. The physics and biology underlying that striking effect remain to be understood yet, in spite of more than 60 years since the concept of FLASH-RT was introduced.\cite{dewey1959}

Recent reviews describe FLASH-RT techniques, implementations, results, and hypothetical mechanisms. \cite{farr2022,schulte2023,limoli2023,chow2024,vozenin2024,rosini2025,feng2025}  The FLASH effect (normal tissue sparing and local tumor control) has been demonstrated with high-energy electron, proton, and x-ray radiation sources for various subjects, from multiple animal models to limited human treatments. Despite many positive findings, a few studies reported the absence of FLASH effect under proton, electron, or photon beams. \cite{vozenin2024} Moreover, while not well documented, anecdotal communications point at cases where once successful, 
the FLASH effect was not observed in the follow up nominally identical experiments, thus calling upon further work to make the modality reliable enough for clinical implementations. From our point of view, the latter outlined observations of 
(i) the FLASH effect itself and (ii) a degree of its elusiveness introduce the essence of FLASH enigma, solving which is necessary to clear a way to further progress. (Some other established facts about FLASH effects will be presented in Section \ref{sec:disc} below.)

At this time, we are not aware of any sufficient answers to the FLASH enigma, although many attempts have been made. Here, we limit ourselves to citing the summary part of a very recent review devoted to the proposed FLASH mechanisms: ``
...oxygen depletion, radical recombination, mitochondrial preservation, DNA repair and immune response modulation, have all been proposed as contributing factors, but no single mechanism fully explains the FLASH effect. '' \cite{rosini2025} 
\begin{figure*}[t!]
\centering
\includegraphics[width = 1.0\textwidth]{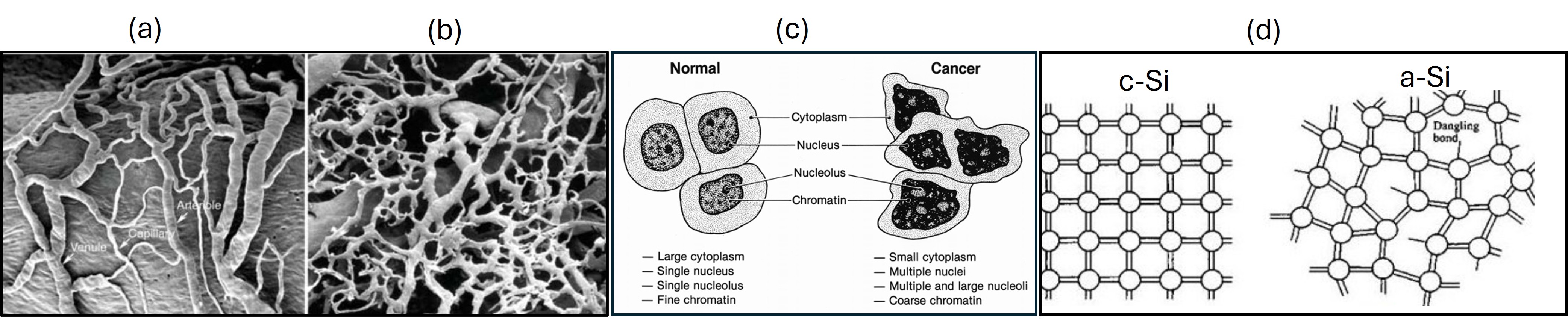}     
\caption{Schematics of ordered vs. disordered counterpart structures. (a) Scanning electron microscopy image of polymer cast of normal microvasculature; (b) same for tumor microvasculature showing disorganization and lack of 
conventional hierarchy of blood vessels. \cite{mcdonald2003} (c) Cell structure in healthy vs. that of cancer tissue. \cite{kenny2001} (d) Crystalline (c-Si) vs. amorphous (a-Si) silicon structure. Note the dangling bonds which can play the role 
of recombination centers. \cite{si}}\label{Fig:disorder}
\end{figure*}

The present paper will take the FLASH problematic down to basic physics. More specifically: 
\begin{itemize}
\item We treat the FLASH enigma as originating in a two-phase system formed by the material of tumor embedded in 
a material of healthy tissues. \item The radiation generates charge carriers (electrons and holes). Upon energy relaxation, they participate in processes creating reactive secondary species (RSS), such are free radicals, essential for destroying cancer cells or healthy cells.
\item The tumor phase exhibits strong structural disorder (heterogeneity) as illustrated in Fig. \ref{Fig:disorder}. \item The disorder
renders anomalously efficient electron-hole recombination and other related pathways allowing efficient energy relaxation. Such efficient pathways exist due to random  structural `defects' (vacancies, dangling bonds, impurities, etc.; a-Si vs. c-Si in Fig. \ref{Fig:disorder} (d) 
is a classical example). Those pathways make the tumor phase `hungry' enough to `digest' all supplied electrons and holes without jamming and produce RSS at any dose rates.   
\cite{mott2012}. \item The healthy tissue is much less disordered not rendering efficient enough energy relaxation pathways. Therefore, under high dose rates, charge carriers accumulate in high concentration forming electron-hole liquid (EHL). A word of caution is in order here: our EHL is defined by the classical condition that energy of interaction between particles exceeds their kinetic energy. The relevance of classical description is guaranteed by a very strong interaction of electrons and holes with surrounding material making them `hydrated' with effective masses on the order of atomic masses (see Sec. \ref{sec:EHLbio}). That classical EHL introduced here should not be confused with the concept of quantum EHL introduced much earlier in connection with crystalline semiconductors \cite{keldysh1970} where effective masses are on the order of electron mass. 
\item Our classical EHL behaves similar to most of the classical molecular liquids with diffusivities much lower than in their corresponding gas phases. The EHL arrested diffusivity suppresses reactions between electrons and holes thus substantially  slowing RSS generation rates, which results in sparing effect.  
\end{itemize}

The ans\"{a}tze of recombination and EHL are additionally outlined in Section \ref{sec:FE} below for the benefits of non-physics readership. For the same, we note that the term of disordered systems does not carry any negative connotation and is broadly used in condensed matter physics. \cite{jain1992,lifshits1988,mott2012}

\section{FLASH enigma}\label{sec:enigma}
We recall that hydroxyl radical (\ce{$^\bullet$}OH) is a commonly recognised RSS important with RT applications. 
It can be generated in a variety of ways \cite{riley1993,vozenin2024,gerin2025} under all types of ionizing radiations.

The radiation generated electrons and holes are initially distributed nonuniformly in random regions (`spurs') of elevated concentrations. Those nonuniformly distributed charge carriers 
diffuse to take part in chemical reactions forming RSS such as hydroxyl radical (\ce{$^\bullet$}OH), etc. The role of diffusion is central here: should the charge carriers be immobile, they would undergo nonradiative recombination returning the system back to its original radical-free state.
We will show how the diffusion of charged species becomes arrested when their concentration grows above a certain value, i. e. a dense conglomerate of charged particles behaves as an electron-hole liquid instead of the conventional low concentration electron hole gas (plasma). The latter understanding is consistent with a general observation that diffusivity drops by 4 to 6 orders of magnitude in the course of phase transformation from gas to liquid. 

The radiation ionized electrons decrease their energy by emitting many low energy atomic vibrations and/or producing secondary ionization events. As a result, they reach the lowest unfilled molecular orbital (LUMO) separated by a significant energy gap $G$ from a positive hole in the highest occupied molecular orbital (HOMO). That gap is significant and the probability of overcoming it is extremely low as explained in Section \ref{sec:rec}. In the meantime, that energy relaxation is necessary to create RSS simply because the energy levels of participating electron and hole must be close to each other to allow a chemical reaction. We will show how it is facilitated in a disordered tumor morphology and hindered in normal tissues under UHDR.

\begin{figure*}[t!]\centering
\includegraphics[width=0.8\textwidth]{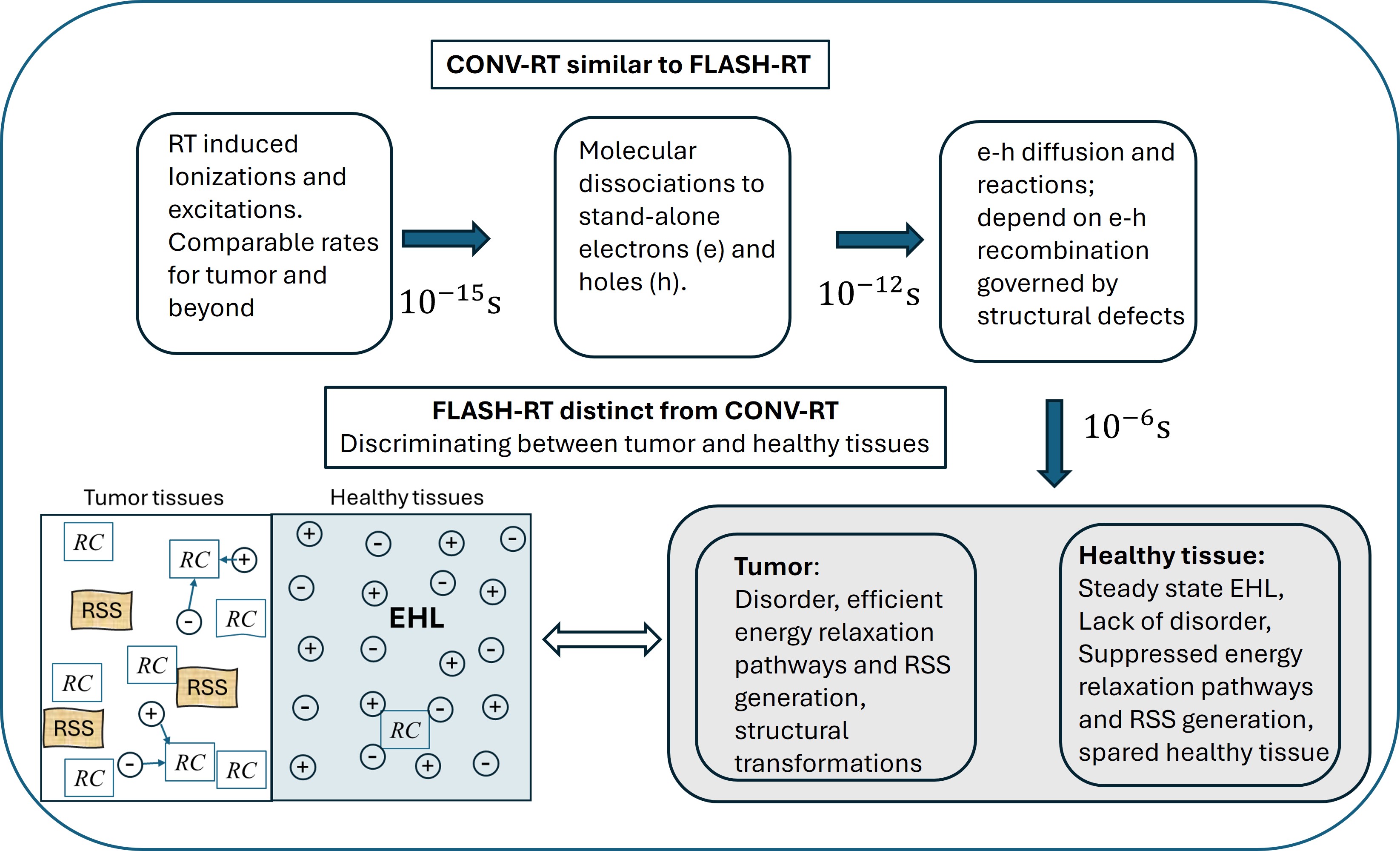}           
\caption{Flow chart aimed at answering the FLASH enigma. The top row boxes (following Fig. 1 of the earlier published review \cite{chow2024}) presents the general physicochemical processes found with both CONV-RT and FLASH-RT 
modalities. The bottom row discriminates between CONV-RT (left) and FLASH-RT (right), as explained in the text. The insert figure in the left bottom corner presents a graphical illustration of the electronic processes under 
irradiation in the tumor (left) and healthy tissues (right) regions. Thin arrows show the e-h recombination processes, $RC$ stand for recombination centers (defects). $+$ and $-$ denote holes and electrons respectively.  }\label{Fig:enigma}
\end{figure*}
Our approach to FLASH enigma is sketched as a process flow in Fig. \ref{Fig:enigma} (cf. Figure 1 in Ref. \cite{chow2024}). 
The top row elements are common between CONV-RT and FLASH-RT. They include ionization and excitation  with 
electrons (e) and holes (h) capable of diffusing and forming RSS. Interaction with 
radiation is practically the same for malignant and normal tissues and so are concentrations of initially generated charge carriers. The subsequent processes of molecular dissociation are almost the same as well because the microscopic molecular structures of those 
phases are similar. 

The rightmost upper box is where the difference between tumor and normal tissues originates.  The CONV-RT regime (left bottom sub-box) presenting here the tumor phase takes place when the charge carriers concentrations 
are not very high and their created stand-alone RSS give rise to DNA transformations. The assumed RSS concentration remains limited even for high dose rates as controlled by a strong energy relaxation/recombination channels 
provided by the high defect concentration characteristic of disordered materials. 
The right bottom sub-box presents the healthy tissue phase where the defect concentration is relatively low and so are the energy relaxation/recombination rates. Given the high dose rate, 
the slow recombination makes the e-h concentration high enough to result in strong Coulomb coupling with energy gain similar to that between the anions and cations in ionic systems, where each charge is bound to the multitude of other charges and cannot be moved. In such a system, 
there are practically no stand-alone charge carriers and corresponding structural transformations are suppressed; hence, spared healthy tissues. The above described scenario is graphically 
illustrated in Fig. \ref{Fig:enigma}.

\section{FLASH electronics}\label{sec:FE}
Here, we briefly discuss the concept of electronic energy relaxation (often referred to as recombination) and that of e-h liquid invoked in the above Sections \ref{sec:intro}, \ref{sec:enigma} which are established in condensed matter physics and yet never mentioned with published FLASH research.

\subsection{e-h recombination}\label{sec:rec} The energy structure of materials at hand can be presented in terms of molecular orbitals LUMO and 
HOMO. Interactions between the molecules transform LUMO and HOMO into conduction and valence bands. \cite{harrison1989} 
The forbidden energy gap $G$ between these bands can be rather wide, say, optical gap $G\sim 10$ eV for the case of water as a prevailing component of biological tissues, \cite{bischoff2021} while the gap for thermal excitations is narrower, $G\approx 6.9$ eV. \cite{shimkevich2014}

Imperfections (lack of periodicity in physical space) result in energy levels inside the energy gap. Possible imperfections include structural defects, such as impurities, dangling bonds, vacancies, etc. Their energy levels represent localized electronic states. In a highly disordered system, the local `defect' structures are random, varying between different locations. Their energy levels found with certain probabilities at any position in the gap. For simplicity, one can  assume a uniform probabilistic distribution of `defect' energy levels in a disordered tumor phase; other, distributions may be invoked if needed \cite{mott2012}. To the contrary, the healthy tissues will have much lower concentration of structural `defects' and their energy levels in the gap. We show next how the latter difference in `defect' energy spectra results in dramatically different recombination rates, high in the tumor and low in the healthy phase.  
\begin{figure}[h!]
\includegraphics[width=0.47\textwidth]{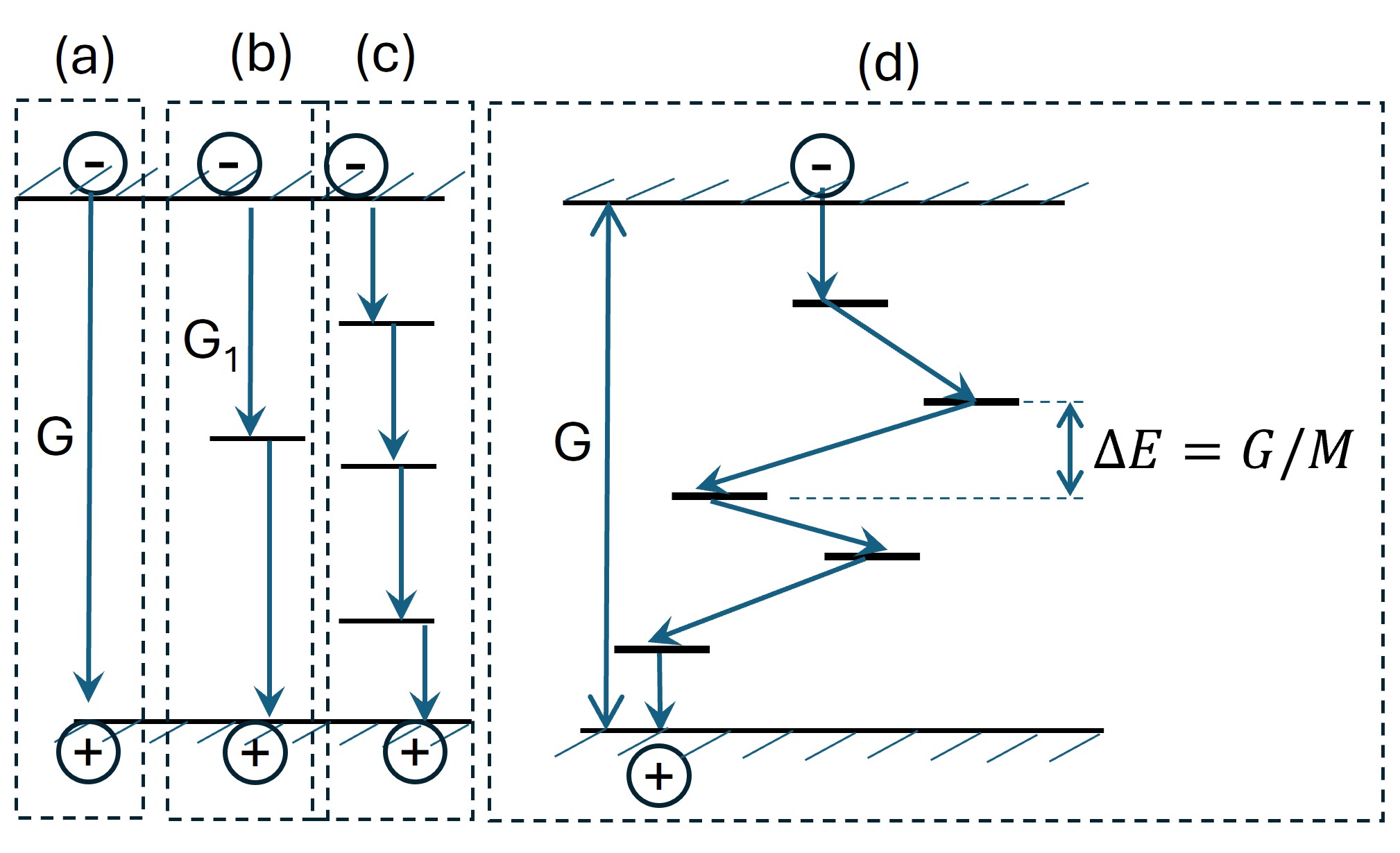}           
\caption{Electronic processes related to FLASH enigma. (a) e-h pair with energy gap $G$ recombining in one step; (b) same pair recombining in two consecutive  steps through a defect state with energy $G_1$; 
(c) multilevel defect states forming a recombination  staircase; (d) same as in (c) in greater detail, showing that each electronic transition takes place not only in energy but in real space as well. Here, vertical axis represents the energy, while the horizontal one corresponds to spatial coordinates. }\label{Fig:electronics}
\end{figure}

Fig. \ref{Fig:electronics} depicts a nonequilibrium (radiation excited) e-h pair. It recombines nonradiatively by dissipating the energy to multiple atomic vibrations (phonons) of relatively low energy, 
say, $\varepsilon\sim 0.03$ eV each. \cite{abakumov1991} Denoting $p_1\ll 1$ the probability of one-phonon emission, the rate of processes emitting $N=G/\varepsilon\gg 1$ phonons is estimated as 
\begin{equation}\label{eq:multph}R_N=p_1^N=\exp\left[-N\ln\left(\frac{1}{p_1}\right)\right]=\exp\left[-\frac{G}{\varepsilon}\ln\frac{1}{p_1}\right]\end{equation}
where the probability logarithm  $\ln(1/p_1)\sim 2-3$.\cite{mott2012} 

Because of the strong inequality $G\gg \varepsilon$, the exponent in Eq. (\ref{eq:multph}) can be significantly reduced by bringing in an additional energy level in the energy gap [shown in Fig. \ref{Fig:electronics} (b)], which takes 
that exponent to $\max[G_1/\varepsilon; (G-G_1)/\varepsilon]$. That reduction is physically related to a two-step recombination process illustrated in Fig. \ref{Fig:electronics} (b). For example, 
it reduces the exponent by a factor of two if we assume $G_1=G/2$, which increases the recombination rate by more than 100 orders of magnitude based on the above mentioned $G=10$ eV $\varepsilon =0.03$ eV, and $\ln(1/p)\sim 3$. 

The latter reasoning shows how the presence of `defects' (impurities, dangling bonds, chemical fluctuations, etc.) exponentially enhances recombination rates. Furthermore, it can be additionally increased by introducing multiple `defect' 
levels so the process evolves in a `staircase', as illustrated in Fig. \ref{Fig:electronics} (c) and (d). A M-step pathway in Fig. \ref{Fig:electronics} (d) will contribute a partial recombination rate $R_M\propto p^M\exp(-G/M\varepsilon )$. The latter product is a sharp maximum when $dR_M/dM=0$. Finding that optimizing $M$ and substituting it back to $R_M$ yields, 
\begin{equation}\label{eq:steps}M=\sqrt{\frac{G}{\varepsilon\ln(1/p)}}, \ R_M\propto \exp\left[-2\sqrt{\frac{G\ln(1/p)}{\varepsilon}}\right] \end{equation}

Eqs. (\ref{eq:multph}) and (\ref{eq:steps}) show that the rates $R_N$ and $R_M$ differ by exponentially large multiplier $\sqrt{G/4\varepsilon}\gg 1$, i.e. the staircase energy relaxation in a disordered system is exponentially more likely than a one-step process with the same energy change in an ordered system. 
More in detail analysis \cite{baranovskii1988,levin1991} accounting for spatial tunneling and polaron effects adds a logarithmic multiplier to the exponent in Eq. (\ref{eq:steps}) retaining the claim of disorder exponentially increasing energy relaxation (`recombination') rates and making them much higher in tumor than in healthy tissues. It predicts exponentially higher RSS generation rates in tumor material, which provides a complementary understanding of the known fact that normal tissues exhibit less radiation susceptibility under UHDR, previously attributed to their self-healing abilities. \cite{limoli2023,chow2024,vozenin2024}

\subsection{Electron hole liquid}\label{sec:EHL}

\begin{figure}[h!]
\includegraphics[width=0.47\textwidth]{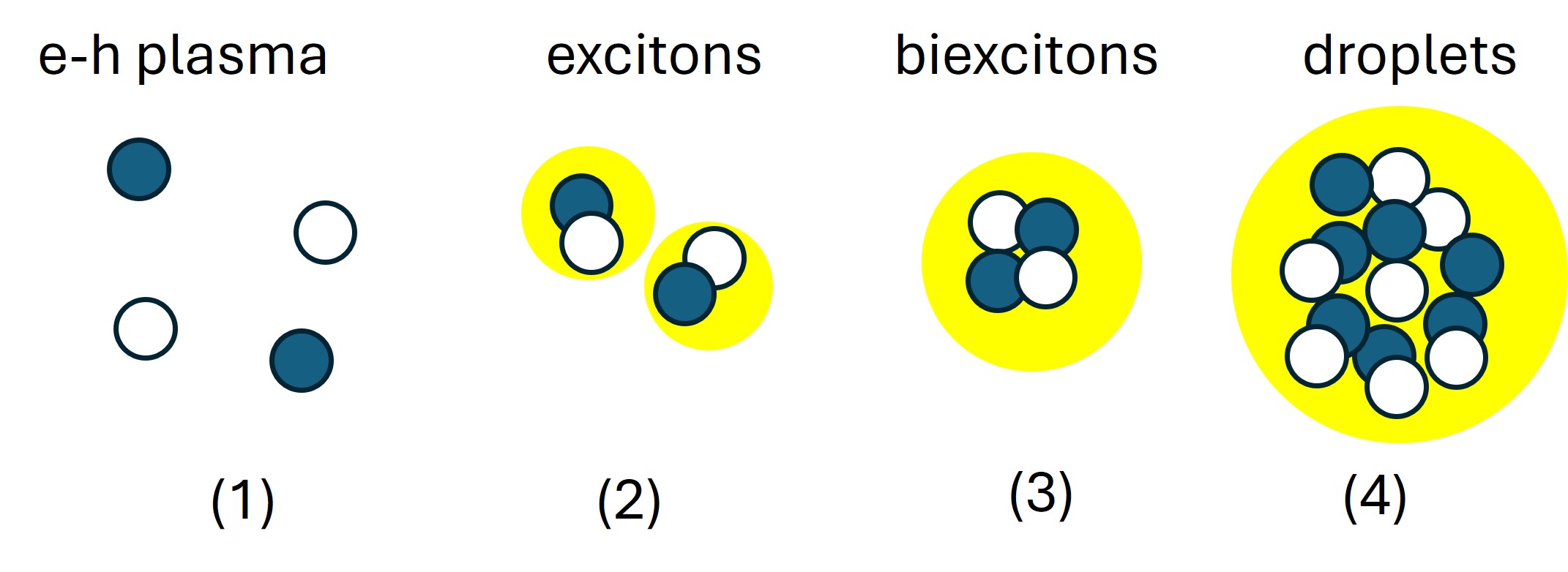}           
\caption{Radiation induced ionization depending on dose rate and energy. Dark circles represent the electrons, white circles represent holes. From right to left: (1) plasma of weakly interacting electrons and holes generated with low dose rates high energy radiation; (2) excitons, i. e. coupled electron -
hole pairs generated with higher dose rates and close to the absorption edge radiation; (3) biexcitons which are coupled pairs of excitons forming with increase of exciton concentration; (4) a droplet of electron hole liquid under high dose rate of well absorbed radiation.}\label{Fig:droplet}
\end{figure}

As illustrated in Fig. \ref{Fig:droplet}, increase in dose rates leads to qualitative changes in the ionized radiation products towards denser e-h conglomerates, through excitons, then biexcitons, towards the electron-hole liquid (EHL). While progress in EHL continues,  \cite{keldysh1970,pokrovskii1972,tikhodeev1985,keldysh1986,ogawa1986,shimano2002,nagai2003,sauer2004,almand2014,poonia2023} many questions remain on both its theoretical and experimental sides. 

The above cited work refers to EHL that is a {\it quantum} phenomenon remaining the only type of EHL known so far; hence, necessary to mention here. The quantum EHL implied a perfect crystalline material where electrons and holes have `normal' effective masses $m$ on the order of bare electron mass and kinetic energies of the order of corresponding band widths, on a scale of electron-volts. To form a bound EHL conglomerate that huge kinetic energy must be counterbalanced by a strong Coulomb interactions between electrons and holes, which makes the problem highly non-trivial. In particular, for such a quantum EHL, there exists the critical temperature $T_c$, below which EHL remains stable; with a few exceptions, $T_c\sim 100$ K. Quantum EHL reveals itself in the observed optical spectra, magnetism, and electrical conduction. For example, it was demonstrated \cite{nagai2003} that EHL photoluminescence emerges hundreds of picosecond later than the stand-alone charge carrier excitations, which reflects the kinetics of EHL condensation from e-h plasma to droplets.

Unlike the {\it quantum} EHL, here introduced {\it classical} EHL pertains to charge carriers with very heavy effective masses on the order of atomic masses, expected in biological systems. Indeed, the charge carriers in biological substances 
strongly interact with relatively soft atomic subsystem creating local atomic rearrangements (polarons, solvated electrons, etc.) in a matter of picoseconds. \cite{emin2013,ghosh2020,bombile2018,conwell2020} Because they carry clouds of atomic deformations, the diffusivities of such charge carriers are extremely low, qualitatively similar to ionic melts. Their kinetic energies are low as well, on the order of $k_BT$ where $k_B$ is the Boltzmann's constant and $T$ is the absolute temperature; $k_BT=0.026$ eV for room temperature $T=300$ K.

Following the standard description of ionic systems, the binding energy in such EHL is given by,
\begin{equation}\label{eq:madelung} U= \frac{\mathfrak{M}e^2}{\kappa r}.\end{equation}
Here, $e$ is the electron charge,  $\kappa$ is the dielectric permittivity, $r$ is the nearest neighbor charge distance, and $\mathfrak{M}$ is the Madelung constant \cite{kittel1996,wikiMad,takenaka2024} accounting for interactions with charges beyond the nearest neighbors. Typically, $\mathfrak{M}\sim 2$ varying between different structures. However, it was argued based on both the data and theoretical analysis that much higher values $\mathfrak{M}\gtrsim 10$ are possible in liquid systems. For definiteness, we will assume $\mathfrak{M}=2$ in our numerical estimates below.
Because of the much lower kinetic energy, the interaction of Eq. (\ref{eq:madelung}) prevails, and the criterion of {\it classical} EHL turns out to be much softer than the above {\it quantum} limitations. It is quantitatively described in Sections {\ref{sec:EHLbio} and \ref{sec:DRC} below.

\subsubsection{Dose criterion for EHL}\label{sec:EHLbio}

To establish the EHL criterion, it is natural to introduce a dimensionless governing parameter, quotient of the interaction and kinetic energy,  
\begin{equation}\label{eq:liquid} L=\frac{U}{k_BT}=\frac{\mathfrak{M}e^2}{\kappa rk_BT}\end{equation}
Depending on its value, a e-h system can behave as a low concentration plasma ($L\ll 1$), or EHL ($L\gtrsim 1$), or ionic crystal ($L\gg 1$). We will see next that $L\gtrsim 1$ for realistic FLASH-RT specifications, i. e. the concept of EHL applies. Interestingly, the EHL concept is qualitatively consistent with the observation of bound pairs between solvated electrons and hydronium cations. \cite{ma2014} 

The binding energy $U$ acts as a diffusion barrier. The probability to overcome it is described by the Boltzmann's exponent, 
\begin{equation}\label{eq:temp}p=\exp(-U/k_BT)=\exp(-L)\end{equation} 
predicting the FLASH effect disappearance with temperature. 
 
To estimate $L$ we start with the integral radiation dose $D$. Assuming that ionized charge carriers do not recombine yields their ultimate concentration, 
\begin{equation}\label{eq:conc}n=2D\rho /I,\end{equation} 
in units 1/cm$^3$ where $\rho \approx 1 $ g/cm$^{3}$ is density of water as the main component, $I$ is the energy of e-h pair creation, and the coefficient 2 accounts for two charge carriers per pair. (Effects of recombination are described next.) The particle concentration and the average distance $r$ between them are related through the standard rigid sphere approximation,
\begin{equation}\label{eq:n-r}n(4\pi r^3/3)=1.\end{equation} 
Substituting here the above $n$ yields
\begin{equation}\label{eq:distance} r=\left(\frac{3I}{8\pi D\rho}\right)^{1/3} \quad {\rm with}\quad D(t)=\int _0^t \dot{D}(t)\, dt\end{equation}
where the dose rate $\dot{D}(t)$ can be time dependent. As a numerical example, the average nearest neighbor distance in EHL is estimated as 7 nm assuming the dose $D=30$ Gy (escalated under the FLASH regime compared to the standard therapeutic values).

Using the latter $r$ in Eq. (\ref{eq:madelung}) yields the value of parameter $L$. Setting then $L>1$ defines a condition on integral dose under which EHL can form,
\begin{equation}\label{eq:flash}D>D_{\rm min}=\left(\frac{\kappa k_BT}{\mathfrak{M}e^2}\right)^3\frac{3I}{8\pi\rho}.\end{equation}

\subsubsection{Recombination and the dose rate criterion for EHL}\label{sec:DRC}
While the criterion of Eq. (\ref{eq:flash}) is given in terms of dose, it is a general perception that the FLASH modality is defined by its dose rate $\dot{D}$. Our resolution here is that the dose of Eq. (\ref{eq:flash}) was derived assuming zero recombination. In reality, a degree of recombination does take place even in healthy phase, which may impose additional restrictions on the FLASH criterion. One relevant restriction is readily established based on qualitative arguments as follows. 

Tacitly assuming zero recombination in deriving Eq. (\ref{eq:flash}) implies that, in the very beginning of radiation exposure, the initial recombination time $\tau (t=0)\equiv \tau _0$ is long enough to allow charge accumulation rather than depletion under charge generation rate $g$ related to the dose rate as 
\begin{equation}\label{eq:genrate}g=\frac{2\rho\dot{D}}{I}.\end{equation} 
Following the standard kinetic analysis we present the recombination driven decrease in carrier concentration as $n=n_0\exp(-t/\tau _0)$, \cite{abakumov1991}. i. e. recombination rate 
\begin{equation}\label{eq:recomb}(\partial n/\partial t)_{\rm recomb}=-n/\tau _0.\end{equation} 
Including both the generation and recombination processes yields, $dn/dt=g-n/\tau _0$. We thus require $dn/dt>0$, i.e. $g>n/\tau _0$. Using the above expression for $g$ and $n=2\rho D/I$, the latter inequality yields the criterion
\begin{equation}\label{eq:doserate} \frac{\dot{D}}{D}>\frac{1}{\tau _0} ,\end{equation}
which should be applied along with the previously derived criterion $D>D_{\rm min}$ of Eq. (\ref{eq:flash}). In a rough approximation, $\dot{D}/D$ expresses a reciprocal of the time $\tau _0$ of dose application; hence, the latter criterion states that the entire dose $D>D_{\rm min}$ must be delivered during a short time not exceeding the initial recombination time in healthy tissues. 

For numerical estimate, we assume $\tau _0\sim 0.01-1$ s from independent study of water electrolysis. \cite{formal2014} Also, we set $I=10$ eV and $\kappa\approx 5$ based on the published data for fat in human tissues \cite{tannino2023} and in bacterial cell. \cite{checa2019} 
Combining the above yields the following criteria for FLASH modality: 
\begin{equation}\label{eq:flash_criteria} D_{\rm min}\sim 10{\rm Gy},\quad \dot{D}_{\rm min}=\frac{D_{\rm min}}{\tau _0}\sim 10-10^3 \frac{{\rm Gy}}{{\rm s}}.\end{equation}
$\tau _0$ can be sensitive to a particular water composition. (With regards to numerical estimates, we note that starting from Eq. (\ref{eq:madelung}) and on we used the Gaussian system expressing $D$ in units erg/g =$10^{-4}$ Gy.)

\subsubsection{Dispersion of dielectric permittivity criterion for EHL}\label{sec:frozen}
Yet another constraint defining the minimum dose rate is related to the time (frequency) dispersion of dielectric permittivity, which, in the low frequency region, is known as $\alpha$ - dispersion. \cite{zimmermann2021} It originates from the relatively slow processes of molecular movements, protein reorientations, etc. determining the low frequency dielectric permittivity of water and related substances. These slow processes make low frequency dielectric permittivity very high, $\kappa\approx 80$ for water. With such dielectric constants, our parameter $L$ in Eq. (\ref{eq:liquid}) decreases by more than order of magnitude which would eliminate the EHL concept. To retain the latter,  $D_{\rm min}$ of Eq. (\ref{eq:flash}) must increase by almost three orders of magnitude than estimated in the above.

However, as pointed out in many cited papers, \cite{zimmermann2021} the slow atomic processes start showing up only at times exceeding 1-10 ms. At shorter time intervals (or frequencies above kHz), they do not contribute, and dielectric permittivity must be mostly of electronic nature, thus about the same as in the above mentioned fat tissues (we note parenthetically that the dielectric permittivity of ice is close to 3.17. \cite{matzler1987}). Coincidentally, the molecular movement constraint time turns out to be of almost the same order of magnitude as our first culprit of recombination, i.e. $\sim 1-10$ ms. Therefore, our estimate in Eq. (\ref{eq:flash_criteria}) remains qualitatively the same. We emphasize here that the `frozen' low frequency molecular dynamic of water appears to be an important part of the FLASH effect in living tissues. The latter contribution of 'frozen' dielectric permittivity is described elsewhere. \cite{shvydka2026}

\subsubsection{Kinetic features}\label{sec:kinf}
Eq. (\ref{eq:recomb}) based approach ignores the EHL caused diffusion suppression. Taking the latter into account dictates renormalization $\tau _0\rightarrow p\tau _0$ with $p$ from Eq. (\ref{eq:temp}) and $r$ expressed through $n$ according to Eq. (\ref{eq:n-r}). As a result, Eq. (\ref{eq:recomb}) takes the form of a kinetic equation,
\begin{equation}\label{eq:kinetics}\frac{dn}{dt}=-\frac{n}{\tau _0}\exp\left[-\frac{\mathfrak{M}e^2}{\kappa k_BT}\left(\frac{4\pi n}{3}\right)^{1/3}\right]+g\end{equation}
where we have added as well the EHL generation rate $g$. 

\begin{figure}[bht] 
\includegraphics[width=0.47\textwidth]{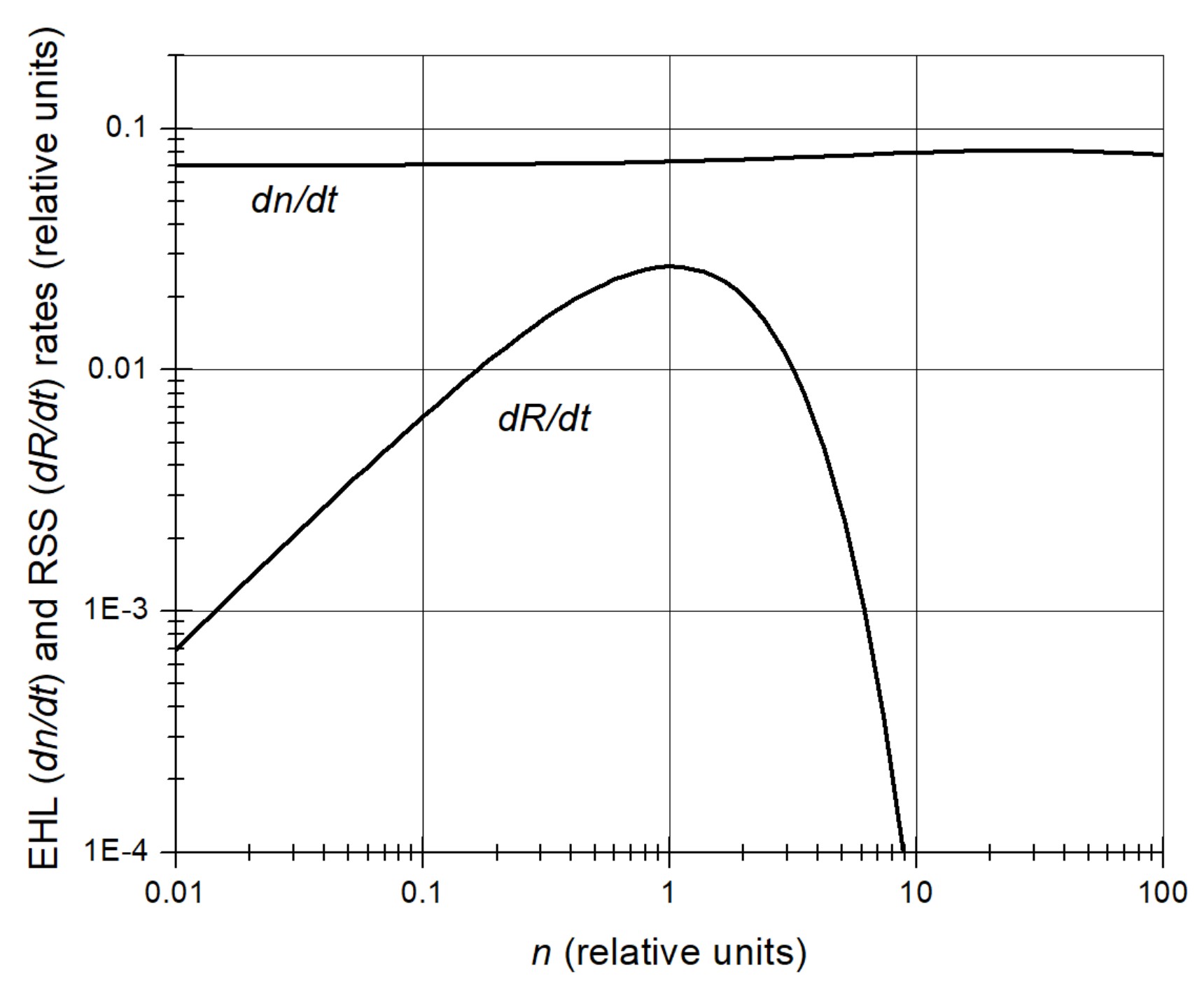} 
\caption{Typical rate dependencies of EHL ($dn(t)/dt$) and RSS ($dR(t)/dt$) on EHL concentration assuming strong enough binding  as described in Eqs. (\ref{eq:kinetics}) and (\ref{eq:RSS}) }\label{Fig:FLASH_rates}
\end{figure}

RSS concentration $R(t)$ generated by charge carriers overcoming the EHL binding is given by,  
\begin{equation}\label{eq:RSS}R(t)=\int _0^t\frac{n}{\tau _0}\exp\left[-\frac{\mathfrak{M}e^2}{\kappa k_BT}\left(\frac{4\pi n}{3}\right)^{1/3}\right]dt.\end{equation}
As illustrated in Fig. \ref{Fig:FLASH_rates} the EHL concentration rate $dn/dt$ slightly increases when the recombination becomes suppressed by the EHL binding. Simultaneously, RSS concentration rate $dR/dt$ riches its maximum and then decreases as the EHL binding becomes strong enough. 

Upon turning the radiation off the EHL binding persists during the time interval of several $\tau _0$ as shown in Fig. \ref{Fig:nt} based on numerical integration of Eq. (\ref{eq:kinetics}) with $g=0$.
\begin{figure}[t!]
\includegraphics[width=0.45\textwidth]{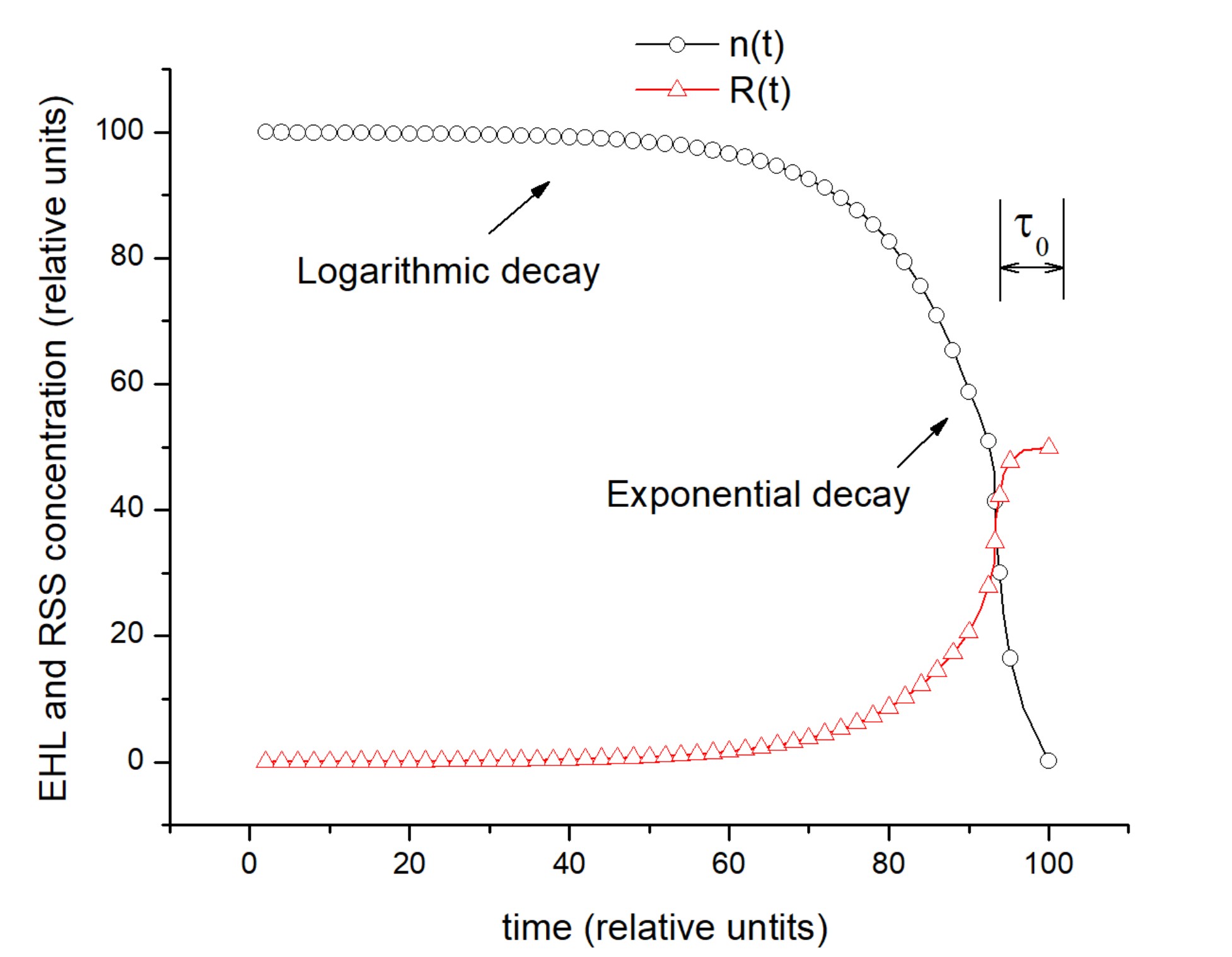}           
\caption{A sketch of temporal dependencies for EHL and RSS concentrations with initial strong binding, $L>2$ upon switching off the radiation source. It was assumed that creation of each RSS particle requires two charge carriers from EHL. }\label{Fig:nt}
\end{figure}

That initial stage of temporal decay in Fig. \ref{Fig:nt} reveals time dependence close to logarithmic. The decay accelerates becoming exponential when EHL concentration decreases enough to weaken the EHL binding. As the binding weakens, electrons and holes acquire the ability to diffuse and engage into chemical reactions creating RSS. Therefore, RSS concentration increases as also depicted in Fig. \ref{Fig:nt}. The underlying modeling procedure is limited to the time domain following radiation shut off (thus not including the EHL and RSS generation), and neglecting RSS diffusion and related destruction processes, not to mention the simplifications of a single recombination time and linear kinetic assumed.  Overall, our modeled `kinetic' features can be rather sensitive to the underlying parameters choice. Our results here are aimed merely at illustrating the major trends. More comprehensive modeling is called upon. 

\subsubsection{Sparing effect}\label{sec:sparing}

Based on Eq. (\ref{eq:kinetics}) we can verify the concept sparing effect (FLASH RT inflicted damage to normal tissue decreasing with dose rate). The damage is associated with the concentration of RSS $R =\int_0^tR(t)dt=$ accumulated during time $t$ and estimated as 
\begin{equation}\label{eq:RSS}R=\int _0^t\frac{n}{\tau _0}\exp\left[-\frac{\mathfrak{M}e^2}{\kappa k_BT}\left(\frac{4\pi n}{3}\right)^{1/3}\right]dt.\end{equation}
we consider short pulses during which the carrier concentration remains unsaturated. Using $n=gt$ and introducing a new variable,
\begin{equation}\label{eq:x}t=\theta x^3\quad {\rm with}\quad \theta =\frac{3}{4\pi g}\left(\frac{\kappa k_BT}{\mathfrak{M}e^2}\right)^3,\end{equation}
one gets
\begin{equation}\label{eq:damage} R =\frac{3g\theta ^2}{\tau _0} \int\limits_{0}^{(t/\theta )^{1/3}} x^5\exp(-x)\,dx  .\end{equation}

In the latter equation, the integrand is a sharp maximum at $x=5$. Therefore, if the upper limit $(t/\theta)^{1/3}$ is considerably greater than 5, the integral is approximately equal $\Gamma (6)=120$, yielding the saturated RSS concentration 
\begin{equation}\label{eq:RSSsat}R_{\rm saturated}=\frac{360g\theta ^2}{\tau _0}=360\left(\frac{D_{\min}\rho}{I}\right)\left(\frac{\dot{D}_{\rm min}}{\dot{D}}\right).\end{equation}
Here, the final expression was written with Eqs.(\ref{eq:flash}), (\ref{eq:genrate}), and (\ref{eq:flash_criteria}) taken into account.

\begin{figure}[b!] 
\includegraphics[width=0.5\textwidth]{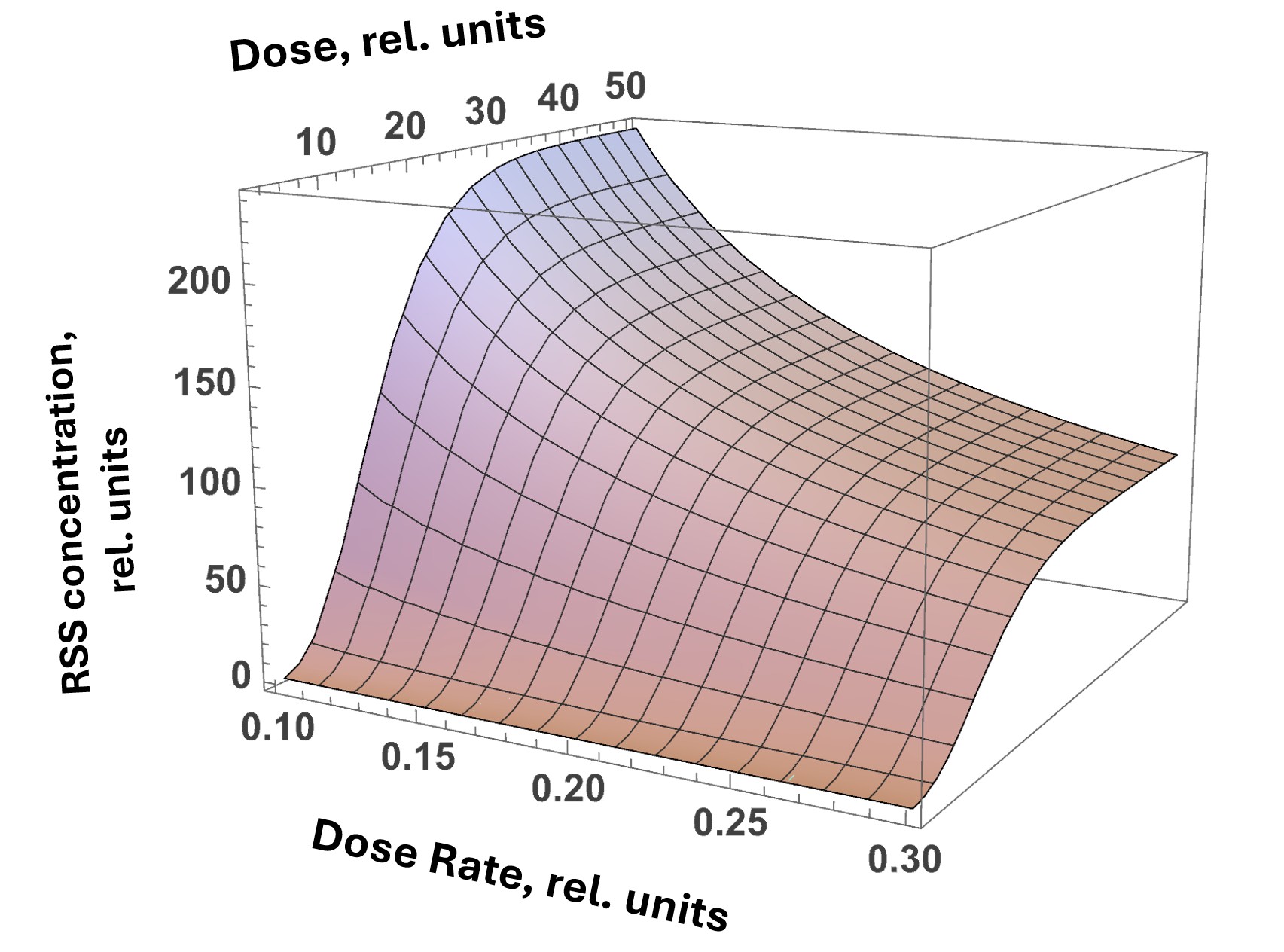} 
\caption{A calculated sketch of sparing effect showing how RT created modifications {\it decrease} with the dose rate for fixed recombination parameters representing normal tissue. }\label{Fig:SPLASH_spare}
\end{figure}

In the opposite limiting case  $(t/\theta)^{1/3}\ll 5$ the integral is estimated as $(1/6)(t/\theta )^2$ yielding the time dependent RSS concentration,
\begin{equation}\label{eq:RSSt}R(t)= \frac{\rho\dot{D}t^2}{3\tau _0I}\end{equation}
where we again used Eqs.(\ref{eq:flash}), (\ref{eq:genrate}).

The results of Eqs. (\ref{eq:RSSsat}) and (\ref{eq:RSSt}) can be traced in the composition 3D plot of Fig.\ref{Fig:SPLASH_spare} presenting $R(t)$ from Eq. (\ref{eq:RSS}) for a broad range of parameters. As seen from that figure and more explicitly from Eq. (\ref{eq:RSSsat}), the sparing effect is reproduced by our theory, attributing it to stronger EHL and its related diffusion arrest under more intense radiation .

We shall end this section by noting that EHL is not expected to exist in the cancer phase where the intrinsic disorder suppresses characteristic recombination times down to $\sim 0.1-1$ $\mu$s, \cite{shen1994}, i.e. several orders of magnitude shorter than the above used $\tau _0\sim 0.01-1$ s. Correspondingly, creating the tissue sparing EHL in cancer phase would require extremely high dose rates $\dot{D}=D_{\rm min}/\tau _0\sim 10^6-10^9$ Gy/s, below which the cancer killing effect prevails.

\subsubsection{Observations consistent with EHL conjecture}\label{sec:someobs}

A recent publication \cite{gupta2023} experimentally evaluated dose rate effects on oxidative damage to peptides, thought of as a simplified model for characterization of the FLASH underlying mechanisms. Utilizing high-energy x-ray and electron sources, the peptides were oxidatively modified depending on dose rate and oxygen availability. It was found that the higher dose rates resulted in less damage. No underlying physical mechanism was determined. Similarly, a more recent publication \cite{debbio2025} explored the molecular basis of the electron FLASH effect in healthy human bronchial epithelial cells with the conclusion that FLASH administration resulted in lower RSS production and consistent reduction of macromolecules’ oxidation, lower cell death and the absence of a cell cycle arrest. We hope that our EHL paradigm can be relevant here explaining how the recombination time $\tau =\tau _0\exp(L)$ increases slowing down the radical supply rate $n/\tau$. One advantage of our explanation is that neither our model nor the experiment \cite{gupta2023,debbio2025} invoke a broad spectrum of biological features often considered for FLASH interpretation (cf. Fig. \ref{Fig:SPLASH_spare}).  

Because the concept of EHL in biological systems is new, no attempts to experimentally observe that state of matter have not yet been performed. However, a substantial effort was made to experimentally observe hydrated electrons under high dose rate radiation in water and aqueous solutions starting from pivotal work several decades ago \cite{keene1960,eiben1962,hart1962,keene1963} and continuing to this day. \cite{kusumoto2024,zhang2024,cao2025} While the FLASH-RT concept had not yet been conceived then, the founding work \cite{keene1960,eiben1962,hart1962,keene1963} coincidentally used the dose rates in the contemporary FLASH domain. That coincidence opens a venue to reattributing some of the existing hydrated electron data \cite{keene1960,eiben1962,hart1962,keene1963,kusumoto2024,cao2025} to EHL.  

Indeed, a strong observation underlying the hydrated electron interpretation  was that the spectral features observed in optical absorption in the irradiated water and aqueous solutions turned out to be independent of specific chemical compositions of a wide range of alkaline substances thus pointing at the intrinsic nature of the observed spectra, reducible to solely water components; hence the hydrated electron. However, EHL appears to be as intrinsic to water as the hydrated electrons, thus rendering an equally satisfying hypothesis. Furthermore, the irradiation induced additional light absorption did not appear with weakly alkaline or acidic solutions. Therefore,  while intrinsic with alkaline substances, the species responsible for the radiation induced additional light absorption disappear when the hydrogen concentration is suppressed.

Finally, we note that due to the general requirement of electric neutrality under irradiation, every hydrated electron must have its positively charged counterpart particle. Regarding of the chemical nature of the latter, they, together with hydrated electrons, will form EHL given the relevant dose rates discussed in Sec. \ref{sec:EHLbio}. 

From a broader perspective, our EHL conjecture introduces a concept of Coulomb correlations applicable to various high ion density electrolytes. Their related observations can be revisited in a quest for such correlations. As our example, we point at the observed change of trend where electrostatic screening length increases with electrolyte concentration above a certain value $n_c$. \cite{smith2024} That trend correlates with EHL effects where the charge carriers lose the ability to screen upon forming strongly correlated conglomerate. Using as a quick estimate the concentration of mobile charges from Eqs. (\ref{eq:kinetics}) and (\ref{eq:RSS}), 
\begin{equation}\label{eq:mobch}n_{mob}=n\exp\left[-\frac{\mathfrak{M}e^2}{\kappa k_BT}\left(\frac{4\pi n}{3}\right)^{1/3}\right]\end{equation}
one can estimate the characteristic concentration $n_c$ via  $dn_{mob}/dn=0$ where $n$ is the total concentration of electrolyte charges, which yields,
\begin{equation}\label{eq:n_c} n_c=\frac{3}{4\pi}\left(\frac{3\kappa k_BT}{\mathfrak{M}e^2}\right)^3,\end{equation}
consistent with the observations \cite{smith2024} at least semiquantitatively. 
Another relevant observation \cite{takenaka2024} may be that of gigantic liquid Madelung energies that can be related to EHL-like correlations in concentrated electrochemical systems. More in detail, the above mentioned observations \cite{smith2024,takenaka2024} will be addressed elsewhere.

\section{Discussion and Conclusions}\label{sec:disc}
Not casting doubts on published biological arguments, our approach is more typical of physics and is based on the ultimate trivialization summarized here. 
The simplest description of the object is a two-phase system consisting of tumor and healthy tissues, along with several intriguing statements related to them.  It was observed that:
\begin{itemize}
\item No matter how high is a dose rate, the tumor phase is able to consume it and its final state is determined by the total dose $D$ absorbed. 
\item A phase of healthy tissues remains perceptive to radiation effects for dose rates below a certain threshold $\dot{D}_c$. However, it locks in becoming irresponsive when dose rate $\dot{D}>\dot{D}_{c}$. 
\end{itemize}

Overall, this paper discriminates between the disordered morphology of tumor vs a more organized structure of normal tissues capable of accumulating high concentration of charge carriers under UHDR. It introduces the concept of EHL attributing sparing of healthy tissues to suppressed chemical reactivity of charge carriers due to EHL binding. Phenomenologically our mechanism of sparing correlates with the "free radical hypothesis" \cite{feng2025} and hypothetical scavenger effects, \cite{alhaddad2024} according to which some undefined processes diminish the concentration of free radicals. Our paper attributes such processes to the conjecture of EHL binding under UHDR radiation. The successes, the remaining questions, and possible experimental verifications of our theory are summarized next.
\subsection{What is understood}\label{sec:wiu}
\begin{flushleft}
(1) Why FLASH exists. The disordered structure of tumor creates efficient energy relaxation pathways consuming most of the electronic excitations under UHDR. On the other hand, more ordered healthy tissue does not provide such pathways making charge carriers accumulate and form EHL that suppresses energy relaxation in the electronic system and slows down RSS formation.\\
(2) Why FLASH has low value thresholds for both the doses and dose rates. Because high enough dose rates are necessary to accumulate e-h concentration sufficient for forming EHL. Our semi-quantitative estimate yields the minimum dose rate of $\sim 10-1000$ Gy/s.\\
(3) Why FLASH (and CONV) RT results depend on dose rather than on dose rate. Because each of the electron-hole pair is utilized by the efficient energy relaxation pathways without jamming in tumor phase.\\
(4) Why healthy tissues are RT sensitive in the sub-threshold (CONV) region. Because EHL is not formed under such dose rates and charge carriers can promptly form RSS. \\
(5) Why healthy tissue in sub-threshold range of dose rates exhibit less sensitivity to RT than cancer. Because their energy transfer pathways are not as efficient as those in cancer phase. (Other explanations are known from biology stating that healthy tissue is capable of self-healing thus exhibiting less damage). \\
(6) A degree of irreproducibility of FLASH effect can be explained by ill-controlled impurities, imperfections and pre-treatments that can destroy EHL in healthy tissues. Such irreproducibility does not show up with CONV RT regime because the latter is not based on the sensitive EHL formation. 
\end{flushleft}
\subsection{What is not understood}\label{sec:winu}
\begin{flushleft}
(1) The quantitative enough temporal kinetic of tumor evolution under FLASH in the framework of above theory.\\
(2) The effects of polaron and/or solvated electron interactions on EHL parameters.\\
(3) Thermal, electromagnetic, and optical responses of EHL as means for experimental verifications.\\
(4) Possible relations between the current theory and biological description. If realistic, to which biological entities, such as, perhaps, free radicals, and/or models would the above described processes correspond?\\
\end{flushleft}
\subsection{Possible experimental verifications}\label{sec:pev}
\begin{flushleft}
(1) Signature properties of EHL created under high dose rates: nontrivial behavior of AC electric conduction, optical absorption, reflection, and photoluminescence during irradiation (when EHL is present in healthy tissue or cell culture).\\
(2) Experimental verification of Eq. (\ref{eq:flash}) for FLASH effect conditions.\\
(3) Spatial variations of RT created effects between various locations in a tumor reflecting its inherent non-uniformity.\\
(4) Doping effects on EHL. It was observed indeed that adding certain impurities can destroy EHL in semiconductors. \cite{ogawa1986} The same could happen with biological EHL, which would destroy the FLASH effect. For example, some protein 
 dopants in healthy tissues could eliminate FLASH. The opposite `positive' effect of doping enhanced FLASH seems theoretically possible. \\
(5) Temperature dependence of the FLASH effect, where decrease in temperature enhances and increase reduces the effect. We note here a recent publication \cite{portier2024} where temperature dependent FLASH effect was observed in some non-cancer cell cultures, but not in those of cancer cells, which is consistent with our EHL interpretation of FLASH. \\
(6) Verification of this paper implication (qualitatively consistent with the observation of Ref. \cite{gupta2023,debbio2025}) that the FLASH (UHDR vs CONV RTfor the same dose/dose rate effect can be observed already at a microscopic level, such as cell cultures, hierarchically below that of organisms. \\ 
(7) Calorimetric verification of EHL phase formation.
\end{flushleft}
\subsection{Special cases and exceptions}\label{sec:sc}
Aiming at conceptual issues, the above analyses are focused on the typical FLASH description: antitumor effectiveness while healthy tissues are spared.  However, some cases of the FLASH dose rates turn out to be less damaging for cancer. \cite{adreian2022}  Taking the latter as an exception, one can attribute it to the energy relaxation decrease caused an atypical distribution of energy levels capable of charge carrier trapping; similar phenomena are known for semiconductors.\cite{bardos2006} On a positive side, the trapping effect may render an opportunity to improve the standard FLASH effect by chemically injecting certain chemical species in healthy tissues prior to treatments. 

As a somewhat related possibility, one can consider the role of interfaces between the cancer and healthy phases. Should the latter be transparent enough, the charge carriers from EHL will penetrate  into the cancer enclosure contributing to its destruction. Effects of that type will depend on the interfacial properties, beyond the present scope.

A broader landscape of special cases can be related to multiple {\it in vitro} studies of FLASH/UHDR effects \cite{adreian2022} relevance to to the corresponding {\it in vivo} cases. Using the standard Epindorphe tubes tissue-culture flasks (Petri-dishes, etc.) brings in a degree of uncertainty because under radiation their forming glass or plastic materials are capable of accumulating significant electric charges thus producing strong electric fields; furthermore, different chemical compositions of the flask materials present different charge polarities. \cite{mizuhara2002,touzin2006,miyake2005,dennison2009,wilson2013} These radiation induced  electric fields \cite{vasko2015} can affect cell cultures. The above effect of radiation induced electrization needs to be taken into account while interpreting the {\it in vitro} FLASH effect data and sub-cellular structures.

\subsection{Summary}\label{sec:sum}

From the practical standpoint, our work points at the principal possibility of testing the FLASH RT parameters at the cellular or sub-cellular level (cell cultures or proteins as opposed to typically stated organismal level) that
significantly simplifies RT planning and verification. In addition, our consideration opens a venue for FLASH RT modifications through various chemical doping schemes affecting recombination and energy relaxation
properties of the living tissues involved. Furthermore, it defines a toolkit of established physical diagnostics, such as electroconductivity, optical absorption and luminescence
predetermining the expected outcomes of FLASH RT. More generally, it brings up the significance of water time/frequency dependent susceptibility ($\alpha$-relaxation etc. that may depend on particular dissolved substances) for nontrivial applications of radiation treatments. Finally, through establishing the existence of high dose rate created electron-hole liquid in water-based substances, our work opens a venue to creating  a technology of high dose rate radiation
detectors needed for FLASH RT applications.

From the academic standpoint, this work presents rather a sketch of theory in its infancy, pointing at important factors and providing order-of-magnitude estimates, yet not
developed enough to quantitatively describe all the details of the FLASH phenomenon. Further effort is called upon to develop this approach. 
    
\section*{acknowledgements} The authors of this manuscript would like to thank Drs. Ahmet Ayan, Ashley Cetnar, Arnab Chakravarti, Eunsin Lee, and Wei Meng, for many enthralling discussions.

\end{document}